\documentclass[12pt]{iopart}

\usepackage{amssymb,amsfonts}
\usepackage{tikz}
\usepackage{color}
\usepackage{ulem,xcolor,mathdots}

\usepackage{epsfig,float,iopams,setstack}
\usepackage{graphicx}

\usepackage{latexsym}
\usepackage{epstopdf}

\begin{document}

\title{Noether symmetries and boundary terms in extended Teleparallel gravity cosmology}
\author{Sebastian Bahamonde}
\ead{sebastian.beltran.14@ucl.ac.uk, sbahamondebeltran@ut.ee}
\address{Laboratory of Theoretical Physics, Institute of Physics, University of Tartu, W. Ostwaldi 1, 50411 Tartu, Estonia}
\address{Department of Mathematics, University College London,
	Gower Street, London, WC1E 6BT, United Kingdom}
\address{School of Mathematics and Physics, University of Lincoln.
	Brayford Pool, Lincoln, LN6 7TS, United Kingdom}
\address{University of Cambridge, Cavendish Laboratory, JJ Thomson Avenue, Cambridge CB3 0HE, United Kingdom}

\author{Ugur Camci}	
\ead{ugurcamci@gmail.com}	
\address{ Siteler Mahallesi, 1307 Sokak, Ahmet Kartal Konutlari, A-1
	Blok, No:7/2, Konyaalti, Antalya, Turkey}

\author{Salvatore Capozziello}
\ead{capozziello@na.infn.it}
\address{Dipartimento di Fisica "E. Pancini", Universit\'a di Napoli
	\textquotedblleft{Federico II}\textquotedblright, Napoli, Italy,}
\address{INFN Sez. di Napoli, Compl. Univ. di Monte S. Angelo, Edificio G, Via
	Cinthia, I-80126,
	Napoli, Italy,}
\address{Laboratory for Theoretical Cosmology, Tomsk State University of Control Systems and Radioelectronics (TUSUR), 634050 Tomsk, Russia.}
	

\begin{abstract}
	We discuss an extended Teleparallel gravity models comprising  functions of scalar invariants constructed by torsion, torsion Gauss-Bonnet and boundary terms. We adopt the Noether Symmetry Approach to select the functional forms, the first integrals and, eventually, the exact solutions of the dynamics in the context of flat Friedman-Robertson-Walker cosmology. Standard perfect fluid matter, minimally coupled to geometry, is considered.  Several physically consistent models are derived with related exact solutions.
	
\end{abstract}


\date{\today}
\maketitle



\section{Introduction}
Cosmological observations such as SNeIa \cite{1}, Baryon Acoustic Oscillations \cite{2} and Cosmic Microwave Background Radiation \cite{3} tell us that our Universe is fuelled by a mysterious ingredient dubbed  dark energy. The role of this entity is accelerating the expansion of the Universe that, otherwise, should  decelerate according to General Relativity (GR) and standard matter (non-exotic) adopted as  sourcing  cosmic fluid.  Then, if GR is assumed to be correct, a new kind of exotic matter, which violates the energy conditions,  is needed to be introduced. On the other hand, if no new material ingredient is detected at fundamental level, GR should be modified or
extended in some way to fit the observed dynamics.
The most well-known model, based on the idea of a further exotic fluid,  is the $\Lambda$CDM where a small cosmological constant $\Lambda$ is the responsible of the acceleration of the Universe \cite{Peebles:2002gy,Carroll:2000fy}. Even though this model fits very well the today cosmological observations \cite{5}, it has some severe theoretical issues such as the so-called cosmological constant problem, the coincidence problem, among others (see~\cite{4,Martin:2012bt,Copeland:2006wr}). Another approach to cure  these theoretical issue is  assuming that GR is valid at Solar system scales and it should be modified at extragalactic and cosmological scales. In other words, GR should be revised both at UV scales (to realize Quantum Gravity) and at IR scales (to fit  cosmological data). This research field is known as modified gravity  and,  in the last 20 years, it has been another approach to  the dark side issues related to both dark energy and dark matter \cite{CapozzielloQuintessence,CapozzielloDark}.  Several alternative  gravity theories have been introduced to fix the phenomenology (see~\cite{Nojiri:2006ri}--\cite{DeFelice:2010aj}  
for some  reviews). The paradigm is that  GR steams out as a particular case of some generalized class of theories   and further degrees of freedom, related to these alternative approaches, are aimed to fit observations without invoking new exotic matter ingredients, up to now not detected at fundamental level.

In this context, Teleparallel equivalent of General Relativity (TEGR) is an alternative and equivalent formulation of GR. This theory relies on a globally flat space-time (zero curvature) with a non-vanishing torsion. The connection of this theory is the so-called Weitzenb\"ock connection $W_{\mu}{}^{a}{}_{\nu}=\partial_{\mu} e^{a}_{\nu}$. In this formalism, tetrads $e^{a}_{\mu}$ are used as the dynamical variables, which are orthonormal vectors at each point of the manifold. Metric can be reconstructed via $g_{\mu\nu}=\eta_{ab}e^{a}_{\mu}e^{b}_{\nu}$, where $\eta_{ab}$ is the Minkowski metric. Torsion tensor is then obtained by taking the skew-symmetric part of the Weitzenb\"ock connection, namely $T^{a}{}_{\mu\nu}=\partial_{\mu}e^{a}_{\nu}-\partial_{\nu}e^{a}_{\mu}$. Note that in this notation, Latin indices refer to the tangent space whereas Greek indices refer to space-time indices. The action of this theory is constructed with the  torsion scalar $T$ recovered by contractions of the torsion tensor $\frac{1}{4} T^\rho{}_{\mu\nu} T_\rho{}^{\mu\nu} + \frac{1}{2}T^\rho{}_{\mu\nu} T^{\nu\mu}{}_{\rho} - T^\lambda{}_{\lambda\mu} T_\nu{}^{\nu\mu}$ and it is directly related to the curvature Ricci scalar $R$ (computed with the Levi-Civita connection) as
\begin{eqnarray}
R=-T+\frac{2}{e}\partial_{\mu}(eT^{\mu})=-T+B\,,\label{RTB}
\end{eqnarray}
where $B$ is a boundary term in the action. Since $R$ and $T$ differs by a boundary term, variations with respect to tetrads give the Einstein field equations. Then, TEGR has the same equations as GR, and, therefore, they give rise to  the same dynamics. However, they are conceptually different physical  theories, for example, in TEGR, geodesic equations are replaced by force equations. For more details about TEGR, see  Refs.~\cite{Obukhov:2002tm}--\cite{Aldrovandi:2013wha}. The above formulation is based on the pure tetrad formalism where it is assumed that the spin connection is zero.

As modifications of GR have been introduced to tackle the cosmological issues, modifications and extensions of  TEGR can be also assumed to address them. They have attained a lot of attention in the last years. The most popular proposal  is $f(T)$ gravity which, instead of assuming a linear combination of $T$ in the action, takes into account a more generic function $f$ depending on $T$  \cite{Ferraro:2006jd,Bengochea:2008gz}. This theory reproduces in very well agreement the current cosmological observations without evoking any dark fluid \cite{Cai:2015emx}--\cite{Aviles:2013nga}. Moreover,  in \cite{Nunes:2018xbm}, it was shown that $f(T)$ would be able to cure the tension for $H_0$. Obviously, this theory is analogous to $f(R)$ gravity~\cite{Capozziello:2011et,DeFelice:2010aj} but they are no longer equivalent due to the boundary contribution $B$. In the pure tetrad formalism, $f(T)$ gravity is not invariant under Lorentz transformations \cite{Cai:2015emx}. This means that two different tetrads, which determine a certain metric, could reproduce different field equations. In \cite{Tamanini:2012hg}, it was shown that one can alleviate this problem by choosing the correct ``good tetrads". Another approach was introduced in \cite{Krssak:2015oua} and further developed in \cite{Krssak:2017nlv,Bahamonde:2017wwk,Golovnev:2017dox}, where a full covariant version of modified TEGR has been studied. Both approaches give the same field equations.

In order to recast these two theories, recently, it was introduced an extension of $f(T)$ gravity which incorporates the boundary term $B$ in the action, the so-called $f(T,B)$ gravity \cite{Bahamonde:2015zma}. Under suitable limits, this theory could become either $f(R)$ or $f(T)$ gravity. Cosmological solutions and properties of this theory have been also discussed in \cite{Bahamonde:2016cul,Bahamonde:2016grb}.  The gravitational energy-momentum pseudo-tensor of $f(T)$ gravity and its relations with the boundary term are discussed in \cite{capriolo}.

In general, this theory is a fourth-order one, so it is more suitable than $f(T)$ to be confronted to $f(R)$ gravity,  which if fourth-order as well. As discussed in \cite{capriolo}, boundary terms determine relations between the two theories also  at the level of gravitational stress-energy pseudo-tensor. In this framework, it is worth considering  also boundary terms  in a comprehensive discussion of dynamics.

An important remark is necessary at this point. In order to consider the whole budget of  degrees of freedom of a given gravity theory, one should consider all the possible geometric invariants at a given order. In \cite{bogdanos}, this issue is considered for fourth-order gravity and then $f(R)$ gravity is extended to $f(R, P,Q)$ where $P\equiv R_{\mu\nu}R^{\mu\nu}$ and
$Q\equiv R_{\alpha\beta\mu\nu}R^{\alpha\beta\mu\nu}$. Such a theory  is equivalent to $f(R, G)$  where $G\equiv R^2-4P+Q$ is the Gauss-Bonnet topological invariant that fixes a constraint among the other curvature invariant. Recently, the Noether symmetries for spatially flat Friedman-Robertson-Walker (FRW) spacetime in Gauss-Bonnet cosmology in $f(R,G)$ gravity  have been studied \cite{capo2014, capo2018,camci2018}, where the authors  used the Noether symmetries as a geometric criterion to select the form of $f(R,G)$ function. It exhausts all the possible degrees of  freedom of  fourth-order curvature gravity (see also \cite{felix}).

A similar approach is possible also in TEGR.
In \cite{Kofinas:2014owa}, it was introduced a  generalized theory where a Teleparallel Gauss-Bonnet term is taken into account. In  curvature approach, besides the above definition, it is well known that in 4-dimensions, the Gauss-Bonnet scalar $G$ is a boundary term. In \cite{Kofinas:2014owa,Bahamonde:2016kba}, it was found that there exists a similar relation,  as~(\ref{RTB}), between the Teleparallel Gauss-Bonnet term $T_G$ and another boundary Gauss-Bonnet term $B_G$, namely
\begin{eqnarray}
G=-T_G+B_G\,,
\end{eqnarray}
where
\begin{eqnarray}
T_{G} &=& (
K_{a}{}^{i}{}_{e}K_{b}{}^{ej}K_{c}{}^{k}{}_{f}K_{d}{}^{fl} -
2K_{a}{}^{ij}K_{b}{}^{k}{}_{e}K_{c}{}^{e}{}_{f}K_{d}{}^{fl} +
2K_{a}{}^{ij}K_{b}{}^{k}{}_{e}K_{f}{}^{el}K_{d}{}^{f}{}_{c}\nonumber\\
&& +
2K_{a}{}^{ij}K_{b}{}^{k}{}_{e}K_{c,d}{}^{el}
)\delta^{abcd}_{ijkl} \,,
\label{eq:defTG}\\
B_G&=&\frac{1}{e}\delta^{abcd}_{ijkl} \partial_{a}
\Big[K_{b}{}^{ij}\Big(K_{c}{}^{kl}{}_{,d}+K_{d}{}^{m}{}_{c}K_{m}{}^{kl}\Big)\Big]\,.
\end{eqnarray}
Here,
\begin{equation}
K_{a}{}^{b}{}_{c}= \frac{1}{2}(T^{\lambda}{}_{\mu\nu} - T_{\nu\mu}{}^{\lambda} +T_{\mu}{}^{\lambda}{}_{\nu})\,,
\end{equation}
is the contorsion tensor, $e={\rm det}(e_{\mu}^a)$,  and $\delta^{abcd}_{ijkl}$ is the generalized Kronecker delta. The two scalars $T_G$ and $B_G$ are also boundary terms in 4-dimensions, therefore they do not contribute to the field equations. However, as long as one considers non-linear terms of these scalars in the action, they contribute to the field equations. Thus, in order to recast curvature Gauss-Bonnet gravity  and Teleparallel Gauss-Bonnet gravity    the following action has been proposed
\cite{Bahamonde:2016kba},
\begin{eqnarray}
S_{f(T,B,T_G,B_G)} = \int
\left[
\frac{1}{\kappa}f(T,B,T_{G},B_{G}) + L_{\rm m}
\right] e\, d^4x \,,\label{action}
\end{eqnarray}
where $\kappa=8\pi G_N$ is the standard gravitational coupling, $L_{\rm m}$ is any matter Lagrangian and now the function $f$ depends on the scalar torsion $T$, the boundary term $B$, the higher-order scalar torsion term $T_G$ and finally the boundary term related to the higher-order term $B_G$. If one chooses $f=f(-T+B,-T_G+B_G)=f(R,G)$,  where $G$ is the above Gauss-Bonnet topological term, the curvature Gauss-Bonnet theory is recovered. The latter theory has been widely discussed in cosmology, see for example \cite{Nojiri:2006ri,Nojiri:2017ncd, cristina, mauro,sergey,micol,alcaniz}. Moreover, if one chooses $f=f(T,T_G)$, one gets Teleparallel Gauss-Bonnet gravity \cite{Kofinas:2014owa} that also has been studied in the context of cosmology in \cite{Kofinas:2014aka,Kofinas:2014daa}. Assuming the gravitational Lagrangian  to depend also on the boundary terms $B_G$ and $B$, the above theory becomes more difficult to analyze. However it  enable us a way to connect all those different theories. The issue of determining the degrees of freedom of Teleparallel theories is an open issue related to the formulation of the theory, its invariances and constraints, see for example~\cite{guzman}.

In this paper, we want to discuss  cosmological solutions coming from the  above theory using the Noether Symmetry Approach \cite{Capozziello:1996bi,Capozziello:2012hm} which is one of the possibilities to find out exact solutions in cosmological context (see also~\cite{kamen}). Using the symmetries of a certain theory, it is possible to constraint the function $f$ in order to find analytical cosmological solutions. This method has been used in $f(T)$ gravity \cite{Atazadeh:2011aa}, $f(T,B)$ gravity \cite{Bahamonde:2016grb}, $f(T,T_G)$ gravity \cite{Capozziello:2016eaz}, scalar-tensor Teleparallel gravity  \cite{Kucukakca:2014vja} and non-local Teleparallel gravity \cite{Bahamonde:2017sdo}. The approach is useful to fix the form of the function $f(T,B,T_{G},B_{G})$, to find cosmological solutions and, eventually, to discuss physically reliable models (see \cite{horndeski} as an example).

This paper is organized as follows. In Sec.~II, the spatially flat FRW cosmology for  $f(T,B,T_G,B_G)$ is introduced. We derive a   point-like Lagrangian by which cosmological equations are derived. Sec.~III is devoted to  the Noether Symmetry Approach and then some exact cosmological solutions are found for different forms of function $f$. Finally, Sec.~IV is devoted to conclusions and outlooks.

\section{ $f(T,B,T_G,B_G)$ cosmology}\label{sec:1}
Let us start our considerations assuming a  flat FRW cosmology given by the tetrad $e^{\mu}_{a}={\rm diag}(1,a(t),a(t),a(t))$ and the metric  $ds^2=-dt^2+a(t)^2(dx^2+dy^2+dz^2)$. If the matter is a perfect fluid with comoving observers $u_{\mu} = \delta^0_{\mu}$, then the energy-momentum tensor is $\mathcal{T}_{\mu \nu} = (\rho + p ) u_{\mu} u_{\nu} + p g_{\mu \nu}$, where $\rho$ is the is the energy density and $p$ is the isotropic pressure measured by the observers $u_{\mu}$. The equation of state is $p = w \rho$ for the perfect fluid, where $w$ is a constant parameter and $w \in [0,1]$ for a barotropic fluid. The parameter describes dust for $w=0$, and a radiation fluid $w= 1/3$. The value $w=-1$ of this parameter corresponds to a cosmological constant. It follows from the conservation law $\mathcal{T}^{\mu \nu}_{\quad ; \nu} = 0$ that $\rho = \rho_{m0} a^{-3 (1 + w)}$, where $\rho_{m0}$ is the energy density of the present universe and $\rho_{m0} = 3 \Omega_{m0} H_0^2$. Starting from the results in \cite{Bahamonde:2016kba}, the cosmological equations are
\begin{eqnarray}
\fl f + \frac{6 \dot{a} \dot{f}_{B}}{a} - \frac{12 f_{T} \dot{a}^2}{a^2} -
\frac{24 \dot{a}^3 \dot{f}_{T_{G}}}{a^3} - \frac{6 f_{B} \left(a \ddot{a}+2 \dot{a}^2\right)}{a^2}+\frac{24 f_{T_{G}} \dot{a}^2 \ddot{a}}{a^3}= \kappa\rho_{m0} a^{-3 (1 + w)}\,,  \label{00} \\
\fl f - \frac{4 \dot{a} \dot{f}_{T}}{a} - \frac{8 \dot{a}^2 \ddot{f}_{T_{G}}}{a^2} -
\frac{6 f_{B} \left(a\ddot{a}+2\dot{a}^2\right)}{a^2} -
\frac{4 f_{T} \left(a\ddot{a}+2 \dot{a}^2\right)}{a^2}-
\frac{16 \dot{a}\ddot{a}\dot{f}_{T_{G}}}{a^2}
\nonumber\\
+\frac{24 f_{T_{G}} \dot{a}^2 \ddot{a}}{a^3}+2 \ddot{f}_{B}
= -\kappa w\rho_{m0} a^{-3 (1 + w)}\,. \label{11}
\end{eqnarray}	
It is important to mention that, for this geometry (flat FRW with $k=0$), the higher-order Gauss-Bonnet term vanishes, i.e. $B_{G}=0$. So, in our case, $f(T,B,T_G,B_G)$ reduces to $f(T,B,T_G)$. Since $G=-T_G+B_G$, the curvature Gauss-Bonnet term and the torsion one are related as $G=-T_G$. Hence, if we choose $f=f(-T+B,-T_G)$, we recover the standard  Gauss-Bonnet cosmology \cite{sergey}. It can be easily found that, in terms of FRW metric,  $T$, $B$ and $T_G$ are given by
\begin{eqnarray}
T &=& 6\Big(\frac{\dot{a}}{a}\Big)^2\,, \label{T} \\
B &=& 6\Big[\frac{\ddot{a}}{a}+2\Big(\frac{\dot{a}}{a}\Big)^2\Big]\,, \label{B} \\
T_G &=& - 24\Big(\frac{\dot{a}}{a}\Big)^2\left(\frac{\ddot{a}}{a}\right)\,. \label{G}
\end{eqnarray}
Let us now rewrite the action (\ref{action}) in a point-like form adopting the FRW metric,  namely
\begin{eqnarray}
& & \fl S_{f(T,B,T_G)} = 2\pi ^{2} \int dt\Big\{ f(T,B, T_G) a^{3} - \lambda_{1} \left[ T - 6 \Big(\frac{\dot{a}}{a}\Big)^2\right]-\lambda_{2} \left( B - 6 \left[\frac{\ddot{a}}{a} + 2\Big(\frac{ \dot{a}}{a}\Big)^2\right]\right) \nonumber\\
&&-\lambda_{3} \left( T_G + 24 \Big(\frac{\dot{a}}{a}\Big)^2\frac{\ddot{a}}{a}\right)- \kappa \rho_{m0} a^{-3 w}\Big\} \,.  \label{actioncan}
\end{eqnarray}
Here $\lambda_1,\lambda_2$ and $\lambda_3$ are the Lagrange multipliers that are easily found by varying this action with respect to $T, B$ and $T_G$, giving us $\lambda_1=a^3f_{T}$, $\lambda_2=a^3f_{B}$ and $\lambda_3=a^3 f_{T_G}$, respectively. Hence, the point-like action reads
\begin{eqnarray}
\fl S_{f(T, B, T_G)} = 2\pi^{2} \int dt\Big\{ a^3 f(T, B, T_G ) - a^3 f_{T} \left[ T - 6 \Big( \frac{\dot{a}}{a} \Big)^2 \right]-a^3 f_{B} \left[ B - 6 \left(\frac{\ddot{a}}{a}+2\Big(\frac{ \dot{a}}{a}\Big)^2\right)\right] \nonumber\\
- a^3 f_{T_G}\left[ T_G  + 24\Big(\frac{\dot{a}}{a}\Big)^2\left(\frac{\ddot{a}}{a}\right) \right] - \kappa \rho_{m0} a^{-3 w}\Big\} \,,  \label{actioncan2}
\end{eqnarray}
which finally gives us the following point-like Lagrangian
\begin{eqnarray}
\fl \mathcal{L}_{f(T, B, T_G)} = 6 a \dot{a}^2 f_{T} - 6 a^2 \dot{a} \dot{f_{B}} + 8 \dot{a}^3 \dot{f}_{T_G} +  a^{3}\Big[ f(T, B, T_G) - T f_T - B f_{B} - T_G f_{T_G} \Big] \nonumber\\
- \kappa \rho_{m0} a^{-3 w}\,,
\label{lagr2}
\end{eqnarray}
where $\dot{f}_B = f_{BT} \dot{T} + f_{BB} \dot{B} + f_{B T_G} \dot{T}_G$ and $\dot{f}_{T_G} = f_{T_G T} \dot{T} + f_{T_G B} \dot{B} + f_{T_G T_G} \dot{T}_G$. Here, we have neglected boundary terms after integration by part. The Lagrangian has a canonical form where kinetic and potential terms can be  clearly distinguished. It is worth saying again that we dealt with the variables $T,  B,$ and $T_G$ under the standard of the Lagrange multipliers. This approach allows us to {\it reduce} dynamics in view to get a canonical Lagrangian considering the constraint Eqs. (\ref{T}), (\ref{B}), and (\ref{G}).
Then we can apply the Noether symmetry approach to the above Lagrangian.
The {\it energy function} associated to $\mathcal{L}$ is defined by
\begin{equation}
E_{\mathcal{L}} = \dot{q}^k \frac{\partial \mathcal{L}}{\partial \dot{q}^k} - \mathcal{L}\,, \label{energy}
\end{equation}
where $q^i (i=1,2,3,4)$ are the generalized coordinates, and $q^i = \{ a, T, B, T_G \}$ for the Lagrangian density (\ref{lagr2}). Thus, the energy functional associated with (\ref{lagr2}) is
\begin{eqnarray}
& & \fl \quad E_{\mathcal{L}} = 6 f_T a \dot{a}^2 - 6 a^2 \dot{a} \dot{f}_B + 24 \dot{a}^3 \dot{f}_{T_G} - a^3 (f - T f_T - B f_B - T_G f_{T_G} ) + \kappa \rho_{m0} a^{-3 w}   \label{energy-2}
\end{eqnarray}
which corresponds to the $00$-component of the field equations, given by (\ref{00}), with the energy condition $E_{\mathcal{L}} = 0$.
The variation of the point-like Lagrangian  (\ref{lagr2}) with respect to the $a, T, B$ and $T_{G}$ gives, respectively
\begin{eqnarray}
& & \fl  f - (T f_T + B f_B + T_G f_{T_G})  - 2 f_T \frac{\dot{a}^2}{a^2} - 4 \left( \frac{ \dot{a} }{a} \dot{f}_T + f_T \frac{\ddot{a}}{a}  \right) - 16 \dot{f}_{T_G} \frac{ \dot{a} \ddot{a} }{a^2} - 8 \ddot{f}_{T_G} \frac{\dot{a}^2}{a^2} + 2 \ddot{f}_B   \nonumber \\ & & \qquad \qquad \qquad \qquad \qquad \qquad \qquad \qquad \qquad \qquad  = - \kappa w \kappa \rho_{m0} a^{-3 w},  \label{feq1} \\& & \fl (6 a \dot{a}^2 - a^3 T) f_{TT} + \left[ 6 ( a^2 \ddot{a} + 2 a \dot{a}^2 ) - a^3 B \right] f_{BT} - \left( 24 \dot{a}^2 \ddot{a} + a^3 T_G \right) f_{T T_G} = 0\,,  \label{feq2} \\& & \fl (6 a \dot{a}^2 - a^3 T) f_{TB} + \left[ 6 ( a^2 \ddot{a} + 2 a \dot{a}^2 ) - a^3 B \right] f_{BB}- \left( 24 \dot{a}^2 \ddot{a} + a^3 T_G \right) f_{B T_G} = 0\,,  \label{feq3}  \\& & \fl (6 a \dot{a}^2 - a^3 T) f_{TG} + \left[ 6 ( a^2 \ddot{a} + 2 a \dot{a}^2 ) - a^3 B \right] f_{BG} - \left( 24 \dot{a}^2 \ddot{a} + a^3 T_G \right) f_{T_G T_G} = 0  \label{feq4}\,,
\end{eqnarray}
which are the Euler-Lagrange equations of motion for the dynamical system given by the
Lagrangian (\ref{lagr2}).  Eq.~(\ref{feq1}) is  the $11$-component of the field equations given by (\ref{11}). It is easily seen for $f_{\alpha \beta} \neq 0$, where $\alpha, \beta \in \{ T, B, T_G \}$, that the equations (\ref{feq2})-(\ref{feq4}) give rise to the expressions (\ref{T})-(\ref{G}) for $T, B$ and $T_G$, and then are compatible with the above  Lagrange multipliers.

\section{Noether's symmetry approach}
Noether's symmetries can give explicit  forms of  Lagrangian (\ref{lagr2}) where cyclic variables, and then conserved quantities exist. This fact allows the exact  integration of  system (\ref{feq1})-(\ref{feq4}) because symmetries are first integrals. The general  approach is outlined below.

Let us consider a Noether symmetry vector generator 
\begin{equation}
{\bf X} = \xi \frac{\partial}{\partial t} + \eta^i \frac{\partial}{\partial q^i}\,, \label{ngs-gen}
\end{equation}
where $q^i = \{ a, T, B, T_G \}$ are the generalized coordinates in the configuration space ${\cal Q }\equiv \{ q^i , i=1,\ldots, 4 \}$ of the Lagrangian, whose tangent space is ${\cal TQ }\equiv \{q^i,\dot{q}^i\}$. The components $\xi$ and $\eta^i$ of the Noether symmetry generator ${\bf X}$ are functions of $r$ and $q^i$. The existence of a Noether symmetry implies the existence of a vector field ${\bf X}$ given in (\ref{ngs-gen}) if the Lagrangian $ \mathcal{L}(t, a, T, B, T_G,  \dot{a}, \dot{T}, \dot{B}, \dot{T}_G )$ satisfies the condition
\begin{equation}
{\bf X}^{[1]} \mathcal{L} + \mathcal{L} ( D_t \xi) = D_t K\,, \label{ngs-eq}
\end{equation}
where ${\bf X}^{[1]}$ is the first prolongation of the generator (\ref{ngs-gen}) in such a form
\begin{equation}
{\bf X}^{[1]} = {\bf X}  + \dot{\eta}^i \frac{\partial}{\partial \dot{q}^i}\,,
\end{equation}
and $K(t, q^i)$ is a gauge function, $D_t$ is the total derivative operator with respect to $t$, $
D_t =\partial / \partial t + \dot{q}^i \partial / \partial q^i$, and $\dot{\eta}^i$ is defined as $\dot{\eta}^i = D_t \eta^i - \dot{q}^i D_t \xi$. It should be noted that both the Lagrangian $\mathcal{L}$ and the prolonged vector field ${\bf X}^{[1]}$ are defined on the first jet bundle, which in this case, is given by $\mathbb{R} \times{\cal TQ }$. It is important to give the following Noether first integral to emphasize the significance of Noether symmetry: if ${\bf X}$ is the Noether symmetry generator corresponding to the Lagrangian $\mathcal{L}(t, q^i, \dot{q}^i)$, then
\begin{equation}
\label{inv}
I =- \xi E_{\mathcal{L}} + \eta^i \frac{\partial \mathcal{L}}{\partial \dot{q}^i} - K \,,
\end{equation}
is also the Hamiltonian or a conserved quantity associated with the generator ${\bf X}$. Now we seek for the condition in order that the Lagrangian density (\ref{lagr2}) would admit a Noether symmetry. The Noether symmetry condition (\ref{ngs-eq}) for the Lagrangian (\ref{lagr2}) gives rise to the following set of differential equations
\begin{eqnarray}
& &\fl \xi_{,a} = 0\,, \quad \xi_{,T} = 0\,, \quad \xi_{,B} = 0\,, \quad \xi_{,T_G} = 0\,, \nonumber \\  & & \fl 12 a f_T \eta^1_{,t} - 6 a^2 \left( f_{BT} \eta^2_{,t} + f_{B B} \eta^3_{,t} + f_{B T_G} \eta^4_{,t}   \right)  - K_{,a} = 0\,,   \nonumber \\  & & \fl  6 a^2 f_{B T} \eta^1_{,t} + K_{,T} = 0\,, \quad  6 a^2 f_{B B} \eta^1_{,t} + K_{,B} = 0\,, \quad 6 a^2 f_{B T_G} \eta^1_{,t} + K_{,T_G} = 0\,,  \nonumber \\  & & \fl f_{T T_G} \eta^1_{,t} = 0\,, \,\, f_{B T_G} \eta^1_{,t} = 0\,, \,\, f_{T_G T_G} \eta^1_{,t} = 0\,, \quad  f_{T T_G} \eta^2_{,t} + f_{B T_G} \eta^3_{,t} + f_{T_G T_G} \eta^4_{,t} = 0\,, \nonumber \\   & & \fl f_{B T} \eta^1_{,T} = 0\,, \quad f_{B B} \eta^1_{,B} = 0\,,\quad f_{B T_G} \eta^1_{,T_G} = 0, \quad f_{T T_G} \eta^1_{,T} = 0,  \nonumber \\   & & \fl f_{B T_G} \eta^1_{,B} = 0\,,  \quad f_{T_G T_G} \eta^1_{,T_G} = 0\,, \quad f_{B T} \eta^1_{,B} + f_{B B} \eta^1_{,T} = 0\,, \nonumber \\ & & \fl  f_{T} \left( \frac{\eta^1}{a} + 2 \eta^1_{,a} - \xi_{,t} \right) + f_{T T} \eta^2 + f_{T B} \eta^3 + f_{T T_G} \eta^4 - a \left( f_{B T_G} \eta^2_{,a} + f_{B B} \eta^3_{,a} + f_{B T_G} \eta^4_{,a}  \right) = 0\,,  \nonumber \\  & & \fl f_{B T} \left( \frac{2 \eta^1}{a} + \eta^1_{,a} + \eta^2_{,T} - \xi_{,t} \right) + f_{B T T} \eta^2 + f_{B B T} \eta^3 + f_{B T T_G} \eta^4 - \frac{2}{a} f_T \eta^1_{,T} \nonumber\\
&& \fl \qquad  + f_{B B} \eta^3_{,T} + f_{B T_G} \eta^4_{,T} = 0\,, \nonumber \\  & & \fl f_{B B} \left( \frac{2 \eta^1}{a} + \eta^1_{,a} + \eta^3_{,B} - \xi_{,t} \right) + f_{B B T} \eta^2 + f_{B B B} \eta^3 + f_{B B T_G} \eta^4 - \frac{2}{a} f_T \eta^1_{,B} \nonumber\\
&& \fl \qquad  + f_{B T} \eta^2_{,B} + f_{B T_G} \eta^4_{,B} = 0\,, \label{ngs} \\  & & \fl  f_{B T_G} \left( \frac{2 \eta^1}{a} + \eta^1_{,a} + \eta^4_{,T_G} - \xi_{,t} \right) + f_{B T T_G} \eta^2 + f_{B B T_G} \eta^3 + f_{B T_G T_G} \eta^4 \nonumber\\
&& \fl  \qquad - \frac{2}{a} f_T \eta^1_{,T_G} + f_{B T_G} \eta^2_{,T_G} + f_{B B} \eta^3_{,T_G} = 0\,, \nonumber \\ && \fl f_{B T} \eta^1_{,T_G} + f_{B T_G} \eta^1_{,T} = 0\,, f_{T T_G} \eta^1_{,B} + f_{B T_G} \eta^1_{,T} = 0\,,f_{B T_G} \eta^1_{,T_G} + f_{T_G T_G} \eta^1_{,B} = 0\,,    \nonumber \\   & & \fl f_{B B} \eta^1_{,T_G} + f_{B T_G} \eta^1_{,B} = 0\,, f_{T T_G} \eta^1_{,T_G} + f_{T_G T_G} \eta^1_{,T} = 0\,,f_{T T_G} \eta^2_{,a} + f_{B T_G} \eta^3_{,a} + f_{T_G T_G} \eta^4_{,a} = 0\,,  \nonumber \\  & & \fl f_{T T_G} \left( 3 \eta^1_{,a} + \eta^2_{,T} - 3 \xi_{,t} \right) + f_{T T T_G} \eta^2 + f_{T B T_G} \eta^3 + f_{T T_G T_G} \eta^4 + f_{B T_G} \eta^3_{,T} + f_{T_G T_G} \eta^4_{,T} = 0\,, \nonumber \\  & & \fl f_{B T_G} \left( 3 \eta^1_{,a} + \eta^3_{,B} - 3 \xi_{,t} \right) + f_{T B T_G} \eta^2 + f_{B B T_G} \eta^3 + f_{B T_G T_G} \eta^4 + f_{T T_G} \eta^2_{,B} + f_{T_G T_G} \eta^4_{,B} = 0\,, \nonumber \\   & & \fl f_{T_G T_G} \left( 3 \eta^1_{,a} + \eta^4_{,T_G} - 3 \xi_{,t} \right) + f_{T T_G T_G} \eta^2 + f_{B T_G T_G} \eta^3 + f_{T_G T_G T_G} \eta^4 \nonumber\\
&& \fl \qquad + f_{T T_G} \eta^2_{,T_G} + f_{B T_G} \eta^3_{,T_G} = 0\,, \nonumber \\   & & \fl V_{,a} \eta^1 + V_{,T} \eta^2 + V_{,B} \eta^3 + V_{,T_G} \eta^4 + V \xi_{,t} + K_{,t} = 0\,, \nonumber
\end{eqnarray}
where $V (a, T, B, T_G) = a^3 \left( T f_T + B f_B + T_G f_{T_G} - f \right) + \kappa \rho_{m0} a^{-3 w}$. Now, assuming compatible  forms of $f(T, B, T_G)$,  we can solve the above system (\ref{ngs}) to derive the components of the symmetry vector ${\bf X}$ given by (\ref{ngs-gen}). Then, we consider the following classes for the form of the function $f(T, B, T_G)$. Here it is pointed out that $\xi = c_1, \eta^i = 0, K=c_2$, where $c_1$ and $c_2$ are arbitrary constants, is a trivial solution of the above system in any case. That is, ${\bf X}_1 = \partial_t$ is a Noether symmetry for any form of the function $f(T, B, T_G)$. This Noether symmetry means that the energy condition $E_{\mathcal{L}} = 0$ has to be satisfied. It is worth saying that the energy functional $E_{\mathcal{L}}$, associated with the Lagrangian,  is defined, in general,  by Eq.(\ref{energy}). From the Noether symmetry equations, it follows that ${\bf X}_1 = \partial_t$ (time translation) is a Noether symmetry which gives the first integral $I = - E_{\mathcal{L}}$ from Eq.(\ref{inv}), where $I$ is a constant of motion. Then, by considering the 00-component of the field equations, we find $I=0$, i.e. $E_{\mathcal{L}} = 0$.
\\ \\
{\bf Case (i):} $f(T, B, T_G) = f_0 T^m + f_1 B^n + f_2 T_G^q$\,.
\\ \\
We find that the unique solution of the above Noether equations (\ref{ngs}) when $f_0,f_1,f_2\neq0$ simultaneously, is for $n = q = 1$. This is trivial because it requires $f(T, B, T_G) = f_0 T^m + f_1 B + f_2 T_G $ which gives rise to the same Noether symmetry as a power-law $f(T)$ function. This comes from the fact that $B$ and $T_G$  are boundary terms so that the linear form of the functions in $B$ and $T_G$ does not introduce any change in the field equations.

In this case, for dust fluid ($w = 0)$, the Noether symmetries for the Lagrangian density (\ref{lagr2}) are obtained as
\begin{eqnarray}
& & \fl \xi = c_1 + c_2 t\,, \quad \eta^1 =c_2 (2 m -1) \frac{a}{3} + c_3 a^{1- \frac{3}{2 m} }\,, \quad \eta^2 = - T \left(  2 c_2 + \frac{3}{m} c_3 a^{-\frac{3}{2 m}} \right)\,, \nonumber \\& & \fl \eta^3 = g (t, a, T, B, T_G), \quad \eta^4 = h (t, a, T, B, T_G), \quad K = - c_2 \kappa \rho_{m0} t + c_4\,, \label{ngsc-1}
\end{eqnarray}
which yields
\begin{eqnarray} \label{ngsv-1}
 \fl {\bf X} &=& (c_1 + c_2 t) \partial_t + \left[ c_2 (2 m -1) \frac{a}{3} + c_3 a^{1 - \frac{3}{2 m}} \right] \partial_a -  \left( 2 c_2 +  \frac{3}{m} c_3 a^{- \frac{3}{2 m}} \right) T \partial_T \nonumber\\
 \fl && +  g \partial_B + h \partial_{T_G}\,,
\end{eqnarray}
where $g$ and $h$ are arbitrary function of $t, a, T, B$ and $T_G$, and $m \neq \frac{1}{2}$. The independent Noether symmetries in this case could be written as ${\bf X}_1 =  \partial_t$ and
\begin{eqnarray} \label{ngsv-1-1}
& &  {\bf X}_2 = t \partial_t +  (2 m -1) \frac{a}{3} \partial_a - 2 T \partial_T  \,\,\, {\rm with} \,\,  K = - \kappa \rho_{m0} t\,, \nonumber\\
&&   {\bf X}_3 =  a^{1 - \frac{3}{2 m}}  \partial_a -  \frac{3}{m} a^{- \frac{3}{2 m}} T \partial_T\,,\quad {\bf X}_4 =  g \partial_B + h \partial_{T_G}\,,
\end{eqnarray}
which have the first integrals
\begin{eqnarray}
&&   I_1 = -  E_{\mathcal{L}}, \quad I_2 =  4 f_0 m (2m -1) T ^{m -1}  a^2  \dot{a} + \kappa \rho_{m0} t\,,\nonumber\\
&&  I_3 = 12 f_0 m T^{m-1} a^{2- \frac{3}{2m}} \dot{a}\,, \quad  I_4 = 0\,, \label{fint-1}
\end{eqnarray}
where $I_1$ vanishes due to the $E_{\mathcal{L}} = 0$. These first integrals has a solution for the scale factor as
\begin{equation}
a(t) = \left[ \frac{I_2 - \kappa \rho_{m0} t }{I_3 (2 m -1)} \right]^{\frac{2m}{3}}\,. \label{a-i-1}
\end{equation}
The Noether symmetry for $m= \frac{1}{2}$ and $w=0$ takes the form
\begin{eqnarray}
& & \fl {\bf X} = \alpha(t) \partial_t + \frac{c_1}{a^2 } \partial_a - 2  \left(  \alpha_{,t} + \frac{3 c_1}{a^3} \right) T \partial_T + g \partial_B + h \partial_{ T_G}  \quad {\rm with} \quad K= - \kappa \rho_{m0} \alpha(t)\,, \label{ngsc-1-4}
\end{eqnarray}
which yields that non-zero first integrals are $ I_1 = \kappa \rho_{m0} \alpha (t)$, i.e., $ \alpha (t) = I_1 / \kappa \rho_{m0}$ is a constant, and
\begin{eqnarray}
& & I_2 = \frac{6 f_0 }{\sqrt{T}} \frac{\dot{a}}{a}\,.  \label{fint-1-4}
\end{eqnarray}
Using the relation (\ref{T}) for $T$, the first integral (\ref{fint-1-4}) yields $I_2 = \sqrt{6} f_0$.

For $w = -1/(2 m)$, the Noether symmetry generator is obtained as follows
\begin{eqnarray}
& & \fl {\bf X} = c_1 \partial_t + \frac{2 m c_2}{3} a^{1 + 3w} \partial_a  - 2 c_2 a^{3 w} T \partial_T + g \partial_B + h \partial_{ T_G} \quad {\rm with} \quad K= c_2 \kappa \rho_{m0} t + c_3\,, \label{ngsc-1-2}
\end{eqnarray}
and the Noether symmetry for $w = 1/(2m-1)$ takes the form
\begin{eqnarray}
& & \fl {\bf X} = (c_1 t + c_2) \partial_t + c_1 (2m -1) \frac{a}{3} \partial_a - 2 c_1 T \partial_T + g \partial_B + h \partial_{ T_G} \quad {\rm with} \quad K = c_3\,,  \label{ngsc-1-3}
\end{eqnarray}
where $c_1, c_2$ and $c_3$ are constant parameters. The corresponding non-zero first-integrals of Noether symmetries (\ref{ngsc-1-2}) and (\ref{ngsc-1-3}) are, respectively,
\begin{eqnarray}
& & I_4 =  8 f_1 m^2 T^{m-1} a^{\frac{4m -3}{2m}} \dot{a} - \kappa \rho_{m0} t\,,  \label{fint-1-2}
\end{eqnarray}
and
\begin{eqnarray}
& & I_5 = 4 f_1 m (2m -1) T^{m-1} a^2 \dot{a}\,.  \label{fint-1-3}
\end{eqnarray}
The solution of the scale factor $a(t)$ from  the latter first integral (\ref{fint-1-3}) is
\begin{equation}
a(t) =  \left[ \frac{2\kappa \rho_{m0}}{3 I_5} m (2m -5) t + a_0 \right]^{-\frac{1}{2m -5}}\,, \label{a-i-3}
\end{equation}
and the former first integral (\ref{fint-1-2}) gives
\begin{equation}
a(t) = a_0 \left( \kappa \rho_{m0} t + I_4 \right)^{\frac{4 m^2}{3 (1- 2m)}}\,, \label{a-i-4}
\end{equation}
where $a_0$ is an integration constant. Using these solutions in the energy condition $E_{\mathcal{L}} = 0$, we find the torsion scalar
\begin{equation}
T = \left[ \frac{\kappa \rho_{m0}}{f_0 (1-2m)} \right]^{1/m} a^{-\frac{6}{2m-1}}\,, \label{T-1}
\end{equation}
for the scale factor (\ref{a-i-3}), and
\begin{equation}
T = \left[ \frac{\kappa \rho_{m0}}{f_0 (1-2m)} \right]^{1/m} a^{\frac{3(1-2m)}{2m^2}}\,, \label{T-2}
\end{equation}
for the scale factor (\ref{a-i-4}). Clearly the dynamical system is completely solved. Some subcases of {\bf Case (i)} are interesting and give rise to further solutions. Let us discuss them.

\bigskip

{\bf Subcase (i-a):} $f(T, B, T_G) =  f_2 T_G^q $.
\\ \\
For this form of the function $f(T, B, T_G)$, the solution of Noether symmetry equations  (\ref{ngs}) of the dust matter ($w = 0$) are
\begin{eqnarray}
& & \xi = c_1 - c_2 t \,, \quad \eta^1 = c_2 (1- 4 q ) \frac{ a}{3}\,, \quad \eta^4 = 4 c_2 T_G\,, \nonumber \\& &  \eta^2 = k (t, a, T, B, T_G)\,, \quad \eta^3 = g (t, a, T, B, T_G)\,, \quad K = c_2 \kappa \rho_{m0} t + c_4\,, \label{ngsc-1-a}
\end{eqnarray}
that is, ${\bf X}_1 = \partial_t$ and
\begin{eqnarray} \label{ngsv-1-a}
& & \fl {\bf X}_2 = t \partial_t + (4 q  -1) \frac{a}{3} \partial_a -  4 T_G \partial_{T_G} \quad  {\rm with} \quad K = - \kappa \rho_{m0} t\,, \quad {\bf X}_3 = k \partial_T + g \partial_B\,,
\end{eqnarray}
are Noether symmetries. Thus, the corresponding first integrals are given by $I_1 = -  E_{\mathcal{L}} = 0, I_3 = 0$ and
\begin{eqnarray}
& & I_2 = 8 f_2 q (q-1) T_G^{q-1}  \left[ (4 q -1) a \dot{a}^2 \frac{\dot{T}_G}{T_G} - 4 \dot{a}^3 \right] + \kappa \rho_{m0} t\,,  \label{fint-1-a}
\end{eqnarray}
where $q\neq 0,1, \frac{1}{4}$. In this subcase, the energy condition $E_{\mathcal{L}} = 0$ gives rise to
\begin{eqnarray}
&& \fl 24 (q-1) \frac{ \dot{a}^3}{a^3} \frac{\dot{T}_G}{T_G^2} + \frac{\kappa \rho_{m0} T_G^{-q} }{q f_2} + 1 = 0\nonumber\\
&& \fl \qquad \Longleftrightarrow  (1- q)  \left( \frac{ \dddot{a}}{\ddot{a}}  - \frac{3 \dot{a}}{a} \right) +  \frac{\ddot{a}}{\dot{a}} \left[ 3 - 2 q + \frac{\kappa \rho_{m0} }{q f_2} \left( -24 \frac{\dot{a}^2 \ddot{a}}{a^3} \right)^{-q} \right] = 0\,. \label{ceq-1}
\end{eqnarray}
It is not found any solution of the above differential equations (\ref{fint-1-a}) and (\ref{ceq-1})  yet. It follows, for $w= 1/(4q -1)$, that the Noether symmetries are ${\bf X}_1, \, {\bf X}_3 $ and
\begin{eqnarray} \label{ngsv-1-a-2}
& & {\bf X}_2 =  w t \partial_t +  \frac{a}{3} \partial_a -  4 w  T_G \partial_{T_G}\,,
\end{eqnarray}
which gives
\begin{eqnarray}
& & I_2 = 8 f_2 q (q-1) T_G^{q-1}  \left[ a \dot{a}^2 \frac{\dot{T}_G}{T_G} - 4 w \dot{a}^3 \right]\,.  \label{fint-1-a-2}
\end{eqnarray}
Here the energy condition $E_{\mathcal{L}} = 0$ yields
\begin{equation}
( q- 1)  \left( \frac{ \dddot{a}}{\ddot{a}}  - \frac{3 \dot{a}}{a} \right) +  ( 2 q -3)  \frac{\ddot{a}}{ \dot{a}} = 0\,,  \label{ceq-1-2}
\end{equation}
and it reduces to
\begin{equation}
\ddot{a} \dot{a}^{\frac{2 q -3}{q-1}}  = a_0 a^3\,,  \label{ceq-1-3}
\end{equation}
where $a_0$ is an integration constant and $q \neq 1, \frac{1}{4}$.
The above equation (\ref{ceq-1-3}) has a solution of the form
\begin{equation}
\left[ 4 (q-1) \right]^{ \frac{q-1}{4 q -5} } \int{ \left[ (4 q -5) a_0 a^4 - a_1  \right]^{ \frac{q-1}{4 q -5} } da  } = t + a_2 \,,  \label{a-1-a}
\end{equation}
where $a_1, a_2$ are constants of integration.

For the function given by $f(T, B, T_G) = f_1 B + f_2 T_G^q $, where $ f_1$ and $ f_2$ are constants, we find the same symmetry (\ref{ngsc-1-a}) of this case. But, for the function $f(T, B, T_G) = f_0 T + f_2 T_G^q $, where $ f_0$ and $ f_2$ are constants, we have only
the Noether symmetry  ${\bf X} = c_1 \partial_t + g(t, a, T, B, T_G) \partial_T + h(t, a, T, B, T_G) \partial_B$, where $g$ and $h$ are arbitrary functions, which gives the energy condition $E_{\mathcal{L}} = 0$, the $00$-component of the field equations.

\bigskip

{\bf Subcase (i-b):} $f(T, B, T_G) = f_1 B^n $.
\\ \\
In this subcase, the Noether symmetries for $w=0$, found from the equations (\ref{ngs}), are ${\bf X}_1$ and
\begin{eqnarray} \label{ngsv-1-b}
& &\fl {\bf X}_2 = t \partial_t + (2 n  -1) \frac{a}{3} \partial_a -  2 B \partial_{B}  \quad {\rm with} \quad K= - \kappa \rho_{m0} t\,, \quad  {\bf X}_3 = k \partial_T + h \partial_{T_G}\,,
\end{eqnarray}
where $k$ and $h$ are arbitrary functions of $t, a, T, B$ and $T_G$. Here the first integrals for ${\bf X}_1, {\bf X}_2$ and ${\bf X}_3$ are $I_1 = -  E_{\mathcal{L}} = 0, I_3 = 0$ and
\begin{eqnarray}
& & I_2 = 2 f_1 n (n-1) B^{n-1}  \left[ 6 a^2 \dot{a}  - (2n -1) a^3 \frac{\dot{B}}{B} \right] + \kappa \rho_{m0} t \,,  \label{fint-1-b}
\end{eqnarray}
which means
\begin{eqnarray}
& &  (1 -2n ) \frac{\dot{B}}{B} + 6  \frac{\dot{a}}{a} = \frac{\left( I_2 - \kappa \rho_{m0} t \right) B^{1-n} }{2 f_1 n (n-1) a^3}\, .  \label{fint-1-b-2}
\end{eqnarray}
where $n \neq 0, 1, \frac{1}{2}$. Now we can consider the energy condition $E_{\mathcal{L}} = 0$ to get
\begin{equation}
6 n \frac{\dot{a} \dot{B}}{ a B} - B = \frac{\kappa \rho_{m0} B^{1-n} }{f_1 (n-1) a^3}\,. \label{ceq-1-b}
\end{equation}
Pulling $B^{1-n}$ from (\ref{ceq-1-b}) and substituting it into (\ref{fint-1-b-2}), one can find that
\begin{equation}
(1 - 2 n) \frac{\dot{B}}{B} + 6 \frac{\dot{a}}{a} +  \frac{( \kappa \rho_{m0} t - I_2)}{2 n \kappa \rho_{m0}} \left( 6 n \frac{\dot{a} \dot{B}}{ a B} - B \right) = 0\,. \label{ceq-1-b-2}
\end{equation}
For $w = 1/(2n -1)$, the Noether symmetries are same as given in the above ${\bf X}_1, {\bf X}_2$ and ${\bf X}_3$ by (\ref{ngsv-1-b}) with vanishing gauge term. Thus the first integral for ${\bf X}_2$ and the energy condition $E_{\mathcal{L}} = 0$ turn out the following form
\begin{eqnarray}
& &  (1 -2n ) \frac{\dot{B}}{B} + 6  \frac{\dot{a}}{a} = \frac{ I_2 }{2 f_1 n (n-1)} \frac{B^{1-n} }{a^3} \,,  \label{fint-1-b-3-1} \\ & & 6 n \frac{\dot{a} \dot{B}}{ a B} - B = \frac{\kappa \rho_{m0} }{f_1 (n-1)} a^{-\frac{6n}{2n -1}}  B^{1-n}\,,   \label{fint-1-b-3-2}
\end{eqnarray}
which yields
\begin{equation}
(1 - 2 n) \frac{\dot{B}}{B} + 6 \frac{\dot{a}}{a} -  \frac{ I_2 a^{ \frac{3}{2n -1}} }{2 n \kappa \rho_{m0}} \left( 6 n \frac{\dot{a} \dot{B}}{ a B} - B \right) = 0\,. \label{ceq-1-b-3}
\end{equation}
For the form of function $f(T, B, T_G) = f_1 B^n + f_2 T_G$, where $f_1$ and $f_2$ are constants, it follows the same Noether symmetries, given by (\ref{ngsv-1-b}) in addition to ${\bf X}_1$, exist. Otherwise, for the function $f(T, B, T_G) = f_0 T + f_1 B^n $, we have the Noether symmetry ${\bf X} = c_1 \partial_t + g(t, a, T, B, T_G) \partial_T + \ell(t, a, T, B, T_G) \partial_{T_G}$, where $g$ and $\ell$ are arbitrary functions. The latter symmetry gives only the energy condition $E_{\mathcal{L}} = 0$ from the first integrals.

\bigskip

{\bf Case (ii):} $f(T, B, T_G) = -T + F(B)$\,.
\\ \\
This case is an extension of TEGR up to a function  depending  on the boundary term. Here $F(B) = B$ is the trivial case which gives rise to the standard TEGR. For $w=0$, we find the following components of the Noether symmetry and the function $F(B)$
\begin{eqnarray}
& & \fl \xi = c_1 t + c_2\,, \quad \eta^1 = \frac{c_1 a}{3} + \frac{c_3 t}{a^2} + \frac{c_4}{a^2}\,, \quad \eta^3 =  - B \left[  2 c_1 + \frac{ 3 (c_3 t + c_4)}{a^3}  \right]\,, \nonumber \\& & \fl \eta^2 = k (t, a, T, B, T_G)\,, \,\, \eta^4 = \ell (t, a, T, B, T_G)\,, \,\, K = -c_1 \kappa \rho_{m0} t  - 2 c_3 \ln \left( B a^3 \right) + c_5\,, \label{ngsc-1-c} \\& & \fl F(B) = f_0 B + \frac{B}{3} \ln B\,, \nonumber
\end{eqnarray}
which means that ${\bf X}_1 = \partial_t $ and
\begin{eqnarray}
& &  {\bf X}_2 = t \partial_t + \frac{a}{3} \partial_a -  2 B \partial_{B} \quad {\rm with} \quad K = - \kappa \rho_{m0} t\,, \nonumber \\& & {\bf X}_3 = \frac{1}{a^2} \partial_a -  \frac{3 B}{a^3} \partial_{B}\,, \quad {\bf X}_4 = t {\bf X}_3 \quad {\rm with } \quad  K = - 2  \ln \left( B a^3 \right)\,,  \label{ngsv-1-c}   \\ & &   {\bf X}_5 = k \partial_T + \ell \partial_{T_G}\,, \nonumber
\end{eqnarray}
are Noether symmetries. Thus, we find the first integrals corresponding to the above Noether symmetries as $I_1 = -  E_{\mathcal{L}}= 0, \, I_5 =0$, and
\begin{eqnarray}
& & \fl \qquad \qquad I_2 =  - \frac{2 a^3}{3} \frac{\dot{B}}{B} + \kappa \rho_{m0} t\,, \quad I_3 = - 2 \left( \frac{\dot{B}}{B} +  \frac{2 \dot{a}}{a} \right)\,, \quad I_4 =  I_3 t + 2 \ln \left( B a^3 \right)\,,  \label{fint-1-c}
\end{eqnarray}
which give the scale factor $a(t)$ and the $B$ term as follows
\begin{eqnarray}
& & a(t) = \left[ a_0 e^{-\frac{I_3 t}{2}} + \frac{3}{I_3} (I_2 - \kappa \rho_{m0} t ) + \frac{6 \kappa \rho_{m0} }{I_3^2} \right]^{\frac{1}{3}}\,, \label{a-i-c} \\ & & B(t) = \frac{e^{\frac{1}{2} (I_4 - I_3 t)}}{a(t)^3}\,,
\end{eqnarray}
where $a_0$ is a constant of integration. This is dust solution ($w=0$) and it is similar of the solution (85) found in \cite{Bahamonde:2016grb}, but our solution explicitly includes Noether integrals of the motion. It is worth noticing that present results are in agreement with those in \cite{Bahamonde:2016grb} but they are richer because we have considered the term $\xi \partial_t$ in the generator which is not considered in \cite{Bahamonde:2016grb}.

For $w=-1$ (the cosmological constant), the Noether symmetries and the function $F(B)$ are obtained by ${\bf X}_1 = \partial_t $ and
\begin{eqnarray}
& &\fl  {\bf X}_2 = t \partial_t + \frac{a}{3} \partial_a -  2 B \partial_{B} \,, \quad {\bf X}_3 = \frac{1}{a^2} \partial_a -  \frac{3 B}{a^3} \partial_{B}\,,  \quad  {\bf X}_4 = t {\bf X}_3 \quad {\rm with } \quad  K = - 2  \ln \left( B a^3 \right)\,,  \nonumber \\ & & \fl  {\bf X}_5 = k \partial_T + \ell \partial_{T_G}, \qquad F(B) = f_0 B + \frac{B}{3} \ln B + \kappa \rho_{m0}\,. \label{ngsv-1-c-2}
\end{eqnarray}
These have the following first integrals
\begin{eqnarray}
& &  I_2 =  - \frac{2 a^3}{3} \frac{\dot{B}}{B}\,, \qquad I_3 = - 2 \left( \frac{\dot{B}}{B} +  \frac{2 \dot{a}}{a} \right)\,, \quad I_4 =  I_3 t + 2 \ln \left( B a^3 \right)\,,  \label{fint-1-c-2}
\end{eqnarray}
which yields
\begin{eqnarray}
& & a(t) = \left[ a_0 e^{-\frac{3 I_3 t}{2}} + \frac{I_2}{I_3} \right]^{\frac{1}{3}}\,, \qquad  B(t) = \frac{e^{\frac{1}{2} (I_4 - I_3 t)}}{a(t)^3}\,. \label{a-i-c-2}
\end{eqnarray}
As far as we know, this is a new cosmological solution.

For $w=1$ (the stiff matter), we find that $F(B) = f_0 B + \frac{B}{3} \ln B $ and the Noether symmetries ${\bf X}_1, {\bf X}_2, {\bf X}_5$ without any gauge term. These have the first integrals $I_1 = - E_{\mathcal{L}} = 0, \, I_2 = -2 a^3 \dot{B} / (3 B)$, and $I_5 = 0$. Using the relation (\ref{B}) for $B$, two of those first integrals have the form
\begin{equation}
\frac{\ddot{a}}{a} - \frac{\dot{a}^2}{a^2} + \frac{3 I_2}{2} \frac{\dot{a}}{a^4} + \frac{\kappa \rho_{m0}}{2 a^6} = 0\,,
\end{equation}
which  gives a solution for the scale factor $a(t)$.

\bigskip

{\bf Case (iii):} $f(T, B, T_G) = f_0 T^m B^n T^q_G$.
\\ \\
The components of Noether symmetry generator ${\bf X} = \partial_t$  for this case are obtained as
\begin{eqnarray}
& &\fl \qquad \xi = c_1 - c_2 t\,, \quad \eta^1 = - c_2 ( 2 m + 2 n + 4 q -1) \frac{a}{3}\,, \quad  \eta^2 = 2 c_2 T\,, \nonumber\\
&&\fl \qquad \eta^3 = 2 c_2 B\,, \qquad \eta^4 = 4 c_2 T_G\,, \qquad K =  c_2 \kappa \rho_{m0} t + c_3\,. \label{ngsc-1}
\end{eqnarray}
Then, the Noether symmetry vectors are ${\bf X}_1 = \partial_t$ and
\begin{eqnarray} \label{ngsv-2}
& & \fl {\bf X}_2 = t \partial_t + (2 m + 2n + 4 q -1) \frac{a}{6} \partial_a -  2 T \partial_T - 2 B \partial_B - 4  T_G \partial_{T_G} \,\,  {\rm with} \,\, K = - \kappa \rho_{m0} t,
\end{eqnarray}
which have the Noether first integrals $I_1 = -  E_{\mathcal{L}} = 0$ and
\begin{eqnarray}
& & \fl  \frac{I_2 - \kappa \rho_{m0} \frac{t}{2} }{ f_0 (2m + 2n + 4 q -1) T^m B^n T_G^q } = m \left( 4 q \frac{a \dot{a}^2}{T_G} - n \frac{a^3}{B}  \right) \frac{\dot{T}}{T} + n \left[ 4 q \frac{a \dot{a}^2}{T_G} - (n-1) \frac{a^3}{B}  \right] \frac{\dot{B}}{B} \nonumber \\& & \fl \qquad \qquad   + q \left[ 4 (q-1) \frac{a \dot{a}^2}{T_G} - n \frac{a^3}{B}  \right] \frac{\dot{T}_G}{T_G}  + \frac{3}{(2m + 2n + 4 q -1)} \left( \frac{4 q}{T_G} \dot{a}^3 - \frac{n}{B} a^2 \dot{a} \right)\,.  \label{fint-2}
\end{eqnarray}
The energy condition $E_{\mathcal{L}} = 0$ for this case becomes
\begin{eqnarray}
& & \fl m \left( \frac{4 q}{T_G} \dot{a}^3 - \frac{n}{B} a^2 \dot{a}   \right) \frac{\dot{T}}{T} +  n \left[ \frac{4 q}{T_G} \dot{a}^3 - \frac{(n-1)}{B} a^2 \dot{a}   \right] \frac{\dot{B}}{B} +  q \left[ \frac{4 (q-1)}{T_G} \dot{a}^3 - \frac{n}{B} a^2 \dot{a}   \right] \frac{\dot{T}_G}{T_G} \nonumber \\& & \fl \qquad \qquad + \frac{m}{T} a \dot{a}^2  + \frac{1}{6} (m + n + q -1) a^3 + \frac{\kappa \rho_{m0}}{6}  = 0\,.
\end{eqnarray}
For the sake of simplicity, hereafter we will study the vacuum case, i.e., $\rho_{m0} = p = 0$ and further assume some subcases.

\bigskip

{\bf Subcase (iii-a):} $f(T, B, T_G) = f_0 T^m B^n $.
\\ \\
The components of Noether symmetry generator (\ref{ngs-gen}) follow from the solution (\ref{ngsc-1}) taking $q=0$, but the component $\eta^4$ is an arbitrary function of $t,a,T,B$ and $T_G$, that is,
\begin{eqnarray}
& &\fl \qquad \xi = c_1 - c_2 t\,, \quad \eta^1 = - c_2 ( 2 m + 2 n - 1) \frac{a}{3}\,, \quad \eta^2 = 2 c_2 T, \nonumber\\
&& \fl \qquad \eta^3 = 2 c_2 B\,, \quad \eta^4 = h(t,a,T,B,T_G)\,, \quad K = c_3\,, \label{ngsc-ii-a}
\end{eqnarray}
which means that the ${\bf X}_1 = \partial_t$ and
\begin{equation}
\fl {\bf X}_2 = t \partial_t + ( 2m + 2n -1) \frac{a}{3} \partial_a - 2 T \partial_T - 2 B \partial_B\,, \quad {\bf X}_3 = h(t,a,T,B,T_G) \partial_{T_G}\,,
\end{equation}
are Noether symmetries. Then, we found the corresponding first integrals as $I_1 = -E_{\mathcal{L}} = 0, \, I_3 = 0$ and
\begin{eqnarray}
\fl \frac{I_2}{f_0 T^m B^n} &= \left[ (2m + 2 n -1) \left( \frac{2 m}{T} + \frac{3 n}{B} \right) - \frac{3 n}{B}  \right] a^2 \dot{a} \nonumber \\ \fl & \qquad  - n (2m + 2 n -1) \frac{a^3}{B} \left[ m \frac{\dot{T}}{T} + (n-1) \frac{\dot{B}}{B} \right]\,. \label{fint-3-1}
\end{eqnarray}
The energy equation $E_{\mathcal{L}} = 0$, in this case, becomes
\begin{equation}
m \frac{\dot{T}}{T} + (n-1) \frac{\dot{B}}{B} = (2 m + n -1) \frac{\dot{a} B}{n a T}\,. \label{energy-3}
\end{equation}
Then, using Eqs. (\ref{energy-3}) in (\ref{fint-3-1}), we obtain
\begin{eqnarray}
& & \frac{I_2}{f_0} T^{1-m} B^{1-n} = 3 n (2m + 2 n -2) a^2 \dot{a} T  - (n -1) (2m + 2 n -1) a^2 \dot{a} B\,. \label{fint-3-2}
\end{eqnarray}
It is interesting to take $n= 1-m$ in the last equation. It gives rise to
\begin{eqnarray}
& & \frac{I_2}{f_0} \left(\frac{B}{T}\right)^{m-1} =  m  a^2 \dot{a}, \quad m \neq 0,1 \label{fint-3-3}
\end{eqnarray}
which  generates  solutions of the scale factor $a(t)$ for any values of $m$.
This case is studied in Ref.~\cite{Bahamonde:2016grb} but taking $m = (1- n)/2$. With this choice of the power of $T$, we find the following Noether symmetries
\begin{eqnarray}
&&  {\bf X}_1 = \partial_t\,,\quad {\bf X}_2 = t \partial_t + n \frac{ a}{3} \partial_a -  2 T  \partial_T - 2 B \partial_B\,, \nonumber\\
&&  {\bf X}_3 = \frac{1}{a^2} \partial_a -  \frac{6 T}{a^3} \partial_T - \frac{3 B}{a^3} \partial_{B}\,, \quad {\bf X}_4 = h(t,a,T,B,T_G) \partial_{T_G}\,, \label{nv-ii-a-1}
\end{eqnarray}
where $h$ is an arbitrary function, and $n\neq 0$. In Ref. \cite{Bahamonde:2016grb}, it is found only the ${\bf X}_3$ as a Noether symmetry. Here, it is explicitly found that there exist  additional Noether symmetry vectors. The first integrals for the Noether symmetries given in (\ref{nv-ii-a-1}) are $I_1 = - E_{\mathcal{L}} = 0, I_4 = 0$, and
\begin{eqnarray}
& & \fl I_2 = 4 f_0 m (1- 2m) T^{m} B^{1-2m} \left(\frac{1}{T} - \frac{3}{B}\right) a^2 \dot{a}\,,  \quad I_3 = 12 f_0 m T^{m-1} B^{1- 2m} \frac{\dot{a}}{a}\,.  \label{fint-3-4}
\end{eqnarray}
Using the constraint $n=1- 2m$ in the energy equation (\ref{energy-3}), it is explicitly seen that $B = \beta \sqrt{T}$, which becomes
\begin{equation}
\frac{\ddot{a}}{a} + \frac{2 \dot{a}^2}{a^2} - \frac{\beta}{\sqrt{6}} \frac{\dot{a}}{a} =0, \end{equation}
where $\beta$ is a constant of integration. This equation has the following solution for the scale factor
\begin{equation}
a(t) =  \left[ a_0 e^{ \frac{\beta t}{\sqrt{6}}} + a_1 \right]^{\frac{1}{3}},  \label{a-ii-a}
\end{equation}
where $a_0$ and $a_1$ are integration constants. This solution is new and different from those found in  Ref.~\cite{Bahamonde:2016grb}. Putting this solution into the first integrals (\ref{fint-3-4}) we have
\begin{equation}
I_2 = \frac{2 \sqrt{6}}{3} f_0 m (1 - 2m) a_1 \beta^{1- 2m}\,, \qquad I_3 = 2 \sqrt{6} f_0 m \beta^{2 m -1}\,. \label{constraints-ii-a}
\end{equation}

\bigskip

{\bf Subcase (iii-b):} $f(T, B, T_G) = f_0 T^m T_G^q $\,.
\\ \\
In this case, taking $n=0$ in (\ref{ngsc-1}), the components of the Noether symmetry take the following form
\begin{eqnarray}
& & \fl \xi = c_1 - c_2 t \,, \quad \eta^1 = - c_2 ( 2 m + 4 q - 1) \frac{a}{3}\,, \quad \eta^2 = 2 c_2 T\,, \quad \eta^4 = 4 c_2 T_G\,,\nonumber\\
&& \fl \eta^3 = g(t,a,T,B,T_G)\,, \quad K = c_3\,. \label{ngsc-4}
\end{eqnarray}
with the  difference that the component $\eta^3$ is an arbitrary function of $t,a,T,B$ and $T_G$.
Then, the Noether symmetries are ${\bf X}_1 = \partial_t$ and
\begin{equation}
\fl {\bf X}_2 = t \partial_t + ( 2m + 4 q - 1) \frac{a}{3} \partial_a - 2 T \partial_T - 4 T_G \partial_{T_G}\,, \quad {\bf X}_3 = g(t,a,T,B,T_G) \partial_{B}\,.
\end{equation}
Here the first integrals for ${\bf X}_1$ and ${\bf X}_3$ vanish, and the first integral for ${\bf X}_2$ is
\begin{eqnarray}
 \fl \frac{I_2}{ 4 f_0 T^m T_G^q} &=&  (2m + 4 q -1) \left[ m \frac{a^2 \dot{a}}{T} + 2 q \frac{a \dot{a}^2}{T_G} \left( \frac{ m \dot{T}}{T} + (q-1)\frac{\dot{T}_G}{T_G} \right) \right] \nonumber\\
\fl &&- 4 q  (m + 2 q - 2) \frac{\dot{a}^3}{T_G}\,.  \label{fint-4}
\end{eqnarray}
The energy condition $E_{\mathcal{L}} = 0$ of this case yields
\begin{equation}
m \frac{\dot{T}}{T} + (q-1) \frac{\dot{T}_G}{T_G} = -\frac{1}{ 4 q}(2 m + q -1) \frac{ a T_G}{\dot{a} T}\,. \label{energy-4}
\end{equation}
Thus, using the above relation in (\ref{fint-4}), it reduce to
\begin{eqnarray}
& & \ell  T^{1-m} T_G^{1-q} =  (1- q) (2m + 4 q -1) a^2 \dot{a} T_G  - 4 q (2m + 4 q -4) \dot{a}^3 T\,, \label{fint-4-2}
\end{eqnarray}
which gives
\begin{eqnarray}
& & \ell T^q T_G^{1- q} = (1- q)(2 q +1) a^2 \dot{a} T_G  + 8 q (1- q) \dot{a}^3 T , \label{fint-4-3}
\end{eqnarray}
for $m= 1- q$,  where $\ell = I_2/(2 f_0)$. The latter equation can also generate some solutions of the scale factor $a(t)$ for any values of $q$. For example, if $q=-1$, then  Eq.~(\ref{fint-4-3}) becomes
\begin{equation}
\dot{a} \ddot{a} - 2 \frac{\dot{a}^3}{a} - 2 \ell \frac{\ddot{a}^2}{a^3} = 0\,. \label{fint-q-1}
\end{equation}
Some solutions of (\ref{fint-q-1}) are
\begin{equation}
a(t) = \left( \frac{ 16 \ell }{ 16 \ell - 3 t} \right)^{\frac{1}{3}}\,, \label{a-4-1}
\end{equation}
for any $\ell$, and
\begin{equation}
a(t) = \sqrt{2} a_0^{1/6} \tan \left( \sqrt{\frac{a_0}{2}} (t + t_0) \right)\,, \label{a-4-2}
\end{equation}
or
\begin{equation}
a(t) = \sqrt{2} a_0^{1/6} \tanh \left( \sqrt{\frac{a_0}{2}} (t + t_0) \right)\,, \label{a-4-3}
\end{equation}
for $I_2 = 1$ and $f_0 = -\frac{1}{2}$, i.e. $\ell = -1$, where $a_0$ is a constant of integration. It is easy to calculate the torsion scalar $T$ and the higher-order scalar torsion Gauss-Bonnet term $T_G$ as
\begin{equation}
T = \frac{5}{(16 \ell - 3 t)^2}\,, \qquad T_G = - \frac{96}{(16 \ell - 3 t)^4}\,
\end{equation}
for the solution (\ref{a-4-1}), and
\begin{equation}
T =  3 a_0 \tan^2 \left( \sqrt{\frac{a_0}{2}} (t + t_0) \right)\,, \qquad T_G = - 12 a_0^2 \frac{\tan^2 \left( \sqrt{\frac{a_0}{2}} (t + t_0) \right) }{ \cos^2 \left( \sqrt{\frac{a_0}{2}} (t + t_0) \right) }\,,
\end{equation}
for the solution (\ref{a-4-2}), and
\begin{equation}
T =  3 a_0 \tanh^2 \left( \sqrt{\frac{a_0}{2}} (t + t_0) \right) \,, \quad T_G =  12 a_0^2 \frac{\tanh^2 \left( \sqrt{\frac{a_0}{2}} (t + t_0) \right) }{ \cosh^2 \left( \sqrt{\frac{a_0}{2}} (t + t_0) \right) }\,,
\end{equation}
for the solution (\ref{a-4-3}). Furthermore, we obtain, from  Eq. (\ref{fint-4-3}) assuming $\ell = 0$, that a power-law solution of the scale factor $a(t)$ is
\begin{equation}
a(t) = a_0 t^{2 q +1}\,, \label{a-4-4}
\end{equation}
which includes a dust-like solution, a radiation-like solution and a stiff matter like solution  for $q= -\frac{1}{6}, -\frac{1}{4}$ and $q= -\frac{1}{3}$, respectively.
Here the torsion scalar $T$ and the higher-order scalar torsion Gauss-Bonnet term $T_G$  behave as $T \sim t^{-2}$ and $T_G \sim t^{-4}$.

Furthermore, Eq.~(\ref{fint-4-3}), for $q= 2$, takes the form
\begin{equation}
80 \dot{a} \ddot{a}^2 - 64 \frac{\dot{a}^3 \ddot{a} }{a} + \ell  = 0\,. \label{fint-q-3}
\end{equation}
This admits the following power law solution
\begin{equation}
a(t) = a_0 \left( t_0 - t \right)^{\frac{5}{3}}\,, \label{a-4-5}
\end{equation}
where $a_0 = 3\left( - 225 \ell \right)^{1/3}/100 $. For this solution, the torsion scalar $T$ and the higher-order scalar torsion Gauss-Bonnet term $T_G$ are found as $T \sim  (t_0 - t)^{-2}$ and $T_G \sim (t_0 - t)^{-4}$.

\bigskip

\section{Conclusions}
In this paper, we have considered an extended Teleparallel gravity where, the function $f(T)$ has been generalized comprising  the Gauss-Bonnet topological invariant and boundary terms. In this perspective, many theories can be recovered from our approach such as curvature Gauss-Bonnet or Teleparallel Gauss-Bonnet gravity. We have not considered non-minimal couplings with scalar fields and other higher-order derivatives of torsion invariants different than the boundary terms $B$ and $B_G$. We have not also included the possible scalars that can be constructed from the decomposition of the torsion tensor $T_{\rm ax},T_{\rm vec}$ and $T_{\rm ten}$ (see~\cite{Bahamonde:2017wwk}). However, for flat FRW cosmology, $T_{\rm ax}=T_{\rm ten}=0$ and $T_{\rm vec}=-9H^2$, so that, flat FRW cosmology for $f(T_{\rm ax},T_{\rm vec},T_{\rm ten})$ will give rise to the same symmetries as $f(T)$ gravity. Therefore, in this space-time, the theory that we consider here, is one of the most general well motivated theories constructed from invariants by torsion tensor.

In order to deal with this dynamics, we adopted the same strategy of the corresponding curvature theory \cite{sergey} searching for the Noether symmetries of dynamical system. In particular, we studied the related FRW cosmology.

Finding out symmetries allows to select the form of $f(T,B,T_G,B_G)$ function, to find out first integrals of motion and, eventually, to find exact solutions. Specifically, in FRW context, it is possible to show that   $f(T,B,T_G,B_G)$  reduces to $f(T,B,T_G)$,  being $B_G=0$.
According to the constraint given in the Noether system (\ref{ngs}), specific forms of $f(T,B,T_G)$ function can be selected.  For any of these functions, Noether vectors can be found and, consequently, first integrals of motions. The process allows to find out exact cosmological solutions for any selected model. Since the solutions have physical meaning, being power law, de Sitter, etc. with a straightforward interpretation, the Noether symmetries correspond to conserved quantities that, eventually, can be directly interpreted. We introduced also minimally coupled standard matter fluid in order to realize more physically consistent systems.

Specifically, conserved quantities related to the existence of Noether symmetries allow to reduce dynamics and then to get  solutions starting from  first integrals. Their physical meaning is related to the fact that, according  the functional form of the Lagrangian, the related exact solutions can be matched with observational data. As discussed in Ref.\cite{perivola}, where scalar-tensor gravity was considered, the functional forms of gravitational coupling and self-interaction potential were derived from the existence of Noether symmetries. The comparison with data, assuming as background the $\Lambda$CDM model, allowed to retain or discard solutions. Physical solutions are considered those matching the observations and capable of  reproducing partially or, in principle totally, the cosmic history.  In the present case,  the reported solutions present power law or exponential behaviors so that, in principle, they can reproduce observed cosmological behaviors.

In a forthcoming paper, the comparison with observational data will be pursued according to the method developed in \cite{perivola}. In this perspective, physically reliable Teleparallel models  can be retained or excluded.

\section*{Acknowledgements}

SB is supported by the Comisi{\'o}n Nacional de Investigaci{\'o}n Cient{\'{\i}}fica y Tecnol{\'o}gica (Becas Chile Grant No.~72150066) and also by Mobilitas Pluss N$^\circ$ MOBJD423 by the Estonian government. SC is supported in part by the INFN sezione di Napoli, {\it iniziativa specifica}  QGSKY. The  article is also based upon work from COST action CA15117 (CANTATA),
supported by COST (European Cooperation in Science and Technology).


\section*{References}


\begin{thebibliography}{99}
	
	
	\bibitem{1}Riess, A.G. et al.: Astron. J. \textbf{116}(1998)1009;Perlmutter, S. et al.: Astrophys. J. \textbf{517}(1999)565.
	
	\bibitem{2}Spergel D.N. et al.: Astrophys. J. Suppl. \textbf{170}(2007)377.
	
	\bibitem{3}Eisenstein, D.J. et al.: Astrophys. J. \textbf{633}(2005)560; E.~Komatsu {\it et al.} [WMAP Collaboration],
	Astrophys.\ J.\ Suppl.\  {\bf 192} (2011) 18
	[arXiv:1001.4538 [astro-ph.CO]]; P.~A.~R.~Ade {\it et al.} [Planck Collaboration],
	Astron.\ Astrophys.\  {\bf 571} (2014) A16
	[arXiv:1303.5076 [astro-ph.CO]].
	
	
	
	\bibitem{Carroll:2000fy}
	S.~M.~Carroll,
	Living Rev.\ Rel.\  {\bf 4} (2001) 1
	[astro-ph/0004075].
	
	
	\bibitem{Peebles:2002gy}
	P.~J.~E.~Peebles and B.~Ratra,
	Rev.\ Mod.\ Phys.\  {\bf 75} (2003) 559
	[astro-ph/0207347].
	
	\bibitem{5}
	M.~Tegmark {\it et al.} [SDSS Collaboration],
	Phys.\ Rev.\ D {\bf 69} (2004) 103501
	[astro-ph/0310723];
	~W.~J.~Percival {\it et al.} [SDSS Collaboration],
	Mon.\ Not.\ Roy.\ Astron.\ Soc.\  {\bf 401} (2010) 2148
	[arXiv:0907.1660 [astro-ph.CO]].
	
	\bibitem{4}Weinberg, S.: Rev. Mod. Phys. \textbf{61}(1989)1.
	
	\bibitem{Martin:2012bt}
	J.~Martin,
	Comptes Rendus Physique {\bf 13}, 566 (2012)
	[arXiv:1205.3365 [astro-ph.CO]].
	\bibitem{Copeland:2006wr}
	E.~J.~Copeland, M.~Sami and S.~Tsujikawa,
	Int.\ J.\ Mod.\ Phys.\ D {\bf 15} (2006) 1753
	[hep-th/0603057].
	\bibitem{CapozzielloQuintessence}
	S.~Capozziello,
	Int.\ J.\ Mod.\ Phys.\ D {\bf 11} (2002) 483
	doi:10.1142/S0218271802002025
	[gr-qc/0201033].
	
	\bibitem{CapozzielloDark}
	S.~Capozziello, V.~F.~Cardone and A.~Troisi,
	JCAP {\bf 0608} (2006) 001
	doi:10.1088/1475-7516/2006/08/001
	[astro-ph/0602349].
	
	
	
	
	
	\bibitem{Nojiri:2006ri}
	S.~Nojiri and S.~D.~Odintsov,
	eConf C {\bf 0602061} (2006) 06
	[Int.\ J.\ Geom.\ Meth.\ Mod.\ Phys.\  {\bf 4} (2007) 115]
	[hep-th/0601213].
	
	\bibitem{Nojiri:2017ncd}
	S.~Nojiri, S.~D.~Odintsov and V.~K.~Oikonomou,
	Phys.\ Rept.\  {\bf 692} (2017) 1
	doi:10.1016/j.physrep.2017.06.001
	[arXiv:1705.11098 [gr-qc]].
	
	
	
	\bibitem{Bamba:2012cp}
	K.~Bamba, S.~Capozziello, S.~Nojiri and S.~D.~Odintsov,
	Astrophys.\ Space Sci.\  {\bf 342} (2012) 155
	doi:10.1007/s10509-012-1181-8
	[arXiv:1205.3421 [gr-qc]].
	
	
	\bibitem{Nojiri:2010wj}
	S.~Nojiri and S.~D.~Odintsov,
	Phys.\ Rept.\  {\bf 505} (2011) 59
	[arXiv:1011.0544 [gr-qc]].
	
	\bibitem{Clifton:2011jh}
	T.~Clifton, P.~G.~Ferreira, A.~Padilla and C.~Skordis,
	Phys.\ Rept.\  {\bf 513} (2012) 1
	[arXiv:1106.2476 [astro-ph.CO]].
	
	
	\bibitem{Capozziello:2011et}
	S.~Capozziello and M.~De Laurentis,
	Phys.\ Rept.\  {\bf 509} (2011) 167
	doi:10.1016/j.physrep.2011.09.003
	[arXiv:1108.6266 [gr-qc]].
	
	\bibitem{DeFelice:2010aj}
	A.~De Felice and S.~Tsujikawa,
	Living Rev.\ Rel.\  {\bf 13} (2010) 3
	doi:10.12942/lrr-2010-3
	[arXiv:1002.4928 [gr-qc]].
	
	\bibitem{Obukhov:2002tm}
	Y.~N.~Obukhov and J.~G.~Pereira,
	Phys.\ Rev.\ D {\bf 67} (2003) 044016
	[gr-qc/0212080].
	
	
	
	\bibitem{Arcos:2005ec}
	H.~I.~Arcos and J.~G.~Pereira,
	Int.\ J.\ Mod.\ Phys.\ D {\bf 13} (2004) 2193
	[gr-qc/0501017].
	
	\bibitem{Weit}
	R.~Weitzenb\"ock,
	{\it Invarianten Theorie}.
	Nordhoff, Groningen (1923).
	
	\bibitem{Cho}
	Y.~M.~Cho,
	Phys.\ Rev.\ D {\bf 14} (1976) 2521.
	
	\bibitem{Cho2}
	Y.~M.~Cho,
	Phys.\ Rev.\ D {\bf 14} (1976) 3335.
	
	\bibitem{Hayashi}
	K.~Hayashi,
	Phys.\ Lett.\ {\bf B69} (1977) 441.
	
	\bibitem{Hay}
	K.~Hayashi and T.~Shirafuji,
	Phys.\ Rev.\ D {\bf 19} (1979) 3524;
	{\bf 24} (1981) 3312.
	
	\bibitem{Kop}
	W. Kopczy\'nski,
	J.\ Phys.\ A {\bf 15} (1982) 493.
	
	\bibitem{Hammond:2002rm}
	R.~T.~Hammond,
	Rept.\ Prog.\ Phys.\  {\bf 65} (2002) 599.
	
	\bibitem{Obukhov1}
	Y. N. Obukhov and G. F. Rubilar, Phys. Rev. D {\bf 73}, 124017 (2006).
	
	\bibitem{Maluf:2013gaa}
	J.~W.~Maluf,
	Annalen Phys.\  {\bf 525} (2013) 339
	[arXiv:1303.3897 [gr-qc]].
	
	\bibitem{Aldrovandi:2013wha}
	R.~Aldrovandi and J.~G.~Pereira,
	{\it Teleparallel Gravity : An Introduction}.
	Fundamental Theories of Physics, Vol. 173.
	Springer Dodrecht, Heidelberg (2013).


	\bibitem{Ferraro:2006jd}
	R.~Ferraro and F.~Fiorini,
	Phys.\ Rev.\ D {\bf 75} (2007) 084031
	[gr-qc/0610067].
	
	\bibitem{Bengochea:2008gz}
	G.~R.~Bengochea and R.~Ferraro,
	Phys.\ Rev.\ D {\bf 79} (2009) 124019
	[arXiv:0812.1205 [astro-ph]].


	
	
	
		
	
	
	\bibitem{Cai:2015emx}
	Y.~F.~Cai, S.~Capozziello, M.~De Laurentis and E.~N.~Saridakis,
	Rept.\ Prog.\ Phys.\  {\bf 79} (2016) no.10,  106901
	doi:10.1088/0034-4885/79/10/106901
	[arXiv:1511.07586 [gr-qc]].
	
	
	
	
	\bibitem{Bamba:2010wb}
	K.~Bamba, C.~Q.~Geng, C.~C.~Lee and L.~W.~Luo,
	JCAP {\bf 1101} (2011) 021
	doi:10.1088/1475-7516/2011/01/021
	[arXiv:1011.0508 [astro-ph.CO]].
	\bibitem{Chen:2010va}
	S.~H.~Chen, J.~B.~Dent, S.~Dutta and E.~N.~Saridakis,
	Phys.\ Rev.\ D {\bf 83} (2011) 023508
	doi:10.1103/PhysRevD.83.023508
	[arXiv:1008.1250 [astro-ph.CO]].
	
	\bibitem{Dent:2011zz}
	J.~B.~Dent, S.~Dutta and E.~N.~Saridakis,
	JCAP {\bf 1101} (2011) 009
	doi:10.1088/1475-7516/2011/01/009
	[arXiv:1010.2215 [astro-ph.CO]].
	
	\bibitem{Bamba:2010wb}
	K.~Bamba, C.~Q.~Geng, C.~C.~Lee and L.~W.~Luo,
	JCAP {\bf 1101} (2011) 021
	doi:10.1088/1475-7516/2011/01/021
	[arXiv:1011.0508 [astro-ph.CO]].
	
	\bibitem{Wu:2010mn}
	P.~Wu and H.~W.~Yu,
	Phys.\ Lett.\ B {\bf 693} (2010) 415
	doi:10.1016/j.physletb.2010.08.073
	[arXiv:1006.0674 [gr-qc]].
	
	\bibitem{Bamba:2012vg}
	K.~Bamba, R.~Myrzakulov, S.~Nojiri and S.~D.~Odintsov,
	Phys.\ Rev.\ D {\bf 85} (2012) 104036
	doi:10.1103/PhysRevD.85.104036
	[arXiv:1202.4057 [gr-qc]].
	
	\bibitem{Wu:2010av}
	P.~Wu and H.~W.~Yu,
	Eur.\ Phys.\ J.\ C {\bf 71} (2011) 1552
	doi:10.1140/epjc/s10052-011-1552-2
	[arXiv:1008.3669 [gr-qc]].
	
	\bibitem{Capozziello:2011hj}
	S.~Capozziello, V.~F.~Cardone, H.~Farajollahi and A.~Ravanpak,
	Phys.\ Rev.\ D {\bf 84} (2011) 043527
	doi:10.1103/PhysRevD.84.043527
	[arXiv:1108.2789 [astro-ph.CO]].
	
	\bibitem{Bengochea:2010sg}
	G.~R.~Bengochea,
	Phys.\ Lett.\ B {\bf 695} (2011) 405
	doi:10.1016/j.physletb.2010.11.064
	[arXiv:1008.3188 [astro-ph.CO]].
	
	\bibitem{Wu:2010xk}
	P.~Wu and H.~W.~Yu,
	Phys.\ Lett.\ B {\bf 692} (2010) 176
	doi:10.1016/j.physletb.2010.07.038
	[arXiv:1007.2348 [astro-ph.CO]].
	
	
	\bibitem{Aviles:2013nga}
	A.~Aviles, A.~Bravetti, S.~Capozziello and O.~Luongo,
	Phys.\ Rev.\ D {\bf 87} (2013) no.6,  064025,
	[arXiv:1302.4871 [gr-qc]]; S.~Capozziello, R.~D'Agostino and O.~Luongo,
	Gen.\ Rel.\ Grav.\  {\bf 49} (2017) no.11,  141,
	[arXiv:1706.02962 [gr-qc]]; S.~Capozziello, O.~Luongo and E.~N.~Saridakis,
	Phys.\ Rev.\ D {\bf 91} (2015) no.12,  124037,
	[arXiv:1503.02832 [gr-qc]].
	
	
	\bibitem{Nunes:2018xbm}
	R.~C.~Nunes,
	arXiv:1802.02281 [gr-qc].
	
	
	
	\bibitem{Tamanini:2012hg}
	N.~Tamanini and C.~G.~Boehmer,
	Phys.\ Rev.\ D {\bf 86} (2012) 044009
	[arXiv:1204.4593 [gr-qc]].
	
	
	\bibitem{Krssak:2015oua}
	M.~Krssak and E.~N.~Saridakis,
	Class.\ Quant.\ Grav.\  {\bf 33} (2016) no.11,  115009
	doi:10.1088/0264-9381/33/11/115009
	[arXiv:1510.08432 [gr-qc]].
	
	\bibitem{Krssak:2017nlv}
	M.~Krssak,
	arXiv:1705.01072 [gr-qc].
	
	\bibitem{Bahamonde:2017wwk}
	S.~Bahamonde, C.~G.~B\"ohmer and M.~Krssak,
	Phys.\ Lett.\ B {\bf 775} (2017) 37
	[arXiv:1706.04920 [gr-qc]].
	
	\bibitem{Golovnev:2017dox}
	A.~Golovnev, T.~Koivisto and M.~Sandstad,
	Class.\ Quant.\ Grav.\  {\bf 34} (2017) no.14,  145013
	[arXiv:1701.06271 [gr-qc]].
	
	\bibitem{Bahamonde:2015zma}
	S.~Bahamonde, C.~G.~B\"ohmer and M.~Wright,
	Phys.\ Rev.\ D {\bf 92} (2015) no.10,  104042
	[arXiv:1508.05120 [gr-qc]].
	
	
	\bibitem{Bahamonde:2016cul}
	S.~Bahamonde, M.~Zubair and G.~Abbas,
	Phys.\ Dark Univ.\  {\bf 19} (2018) 78
	[arXiv:1609.08373 [gr-qc]].
	
	\bibitem{Bahamonde:2016grb}
	S.~Bahamonde and S.~Capozziello,
	Eur.\ Phys.\ J.\ C {\bf 77} (2017) no.2,  107
	[arXiv:1612.01299 [gr-qc]].
	
	
	\bibitem{capriolo}
	S.~Capozziello, M.~Capriolo and M.~Transirico,
	arXiv:1804.08530 [gr-qc], to appear in Int. Jou. Geom. Meth.  Mod. Phys. (2018).
	
	
	\bibitem{bogdanos}
	C.~Bogdanos, S.~Capozziello, M.~De Laurentis and S.~Nesseris,
	Astropart.\ Phys.\  {\bf 34} (2010) 236
	[arXiv:0911.3094 [gr-qc]].

    \bibitem{capo2014} S. Capozziello, M. De Laurentis and S. D. Odintsov, 
     Mod.\ Phys.\ Lett. A \ {\bf 29} (2014) 1450164 [ArXiv:1406.5652 [gr-qc]].
    
    \bibitem{capo2018} K. F. Dialektopoulos and S. Capozziello, 
     Int.\ J.\ Geom.\ Meth.\ Mod.\ Phys.\ {\bf 15} (2018) 1840007 [ArXiv:1808.03484 [gr-qc]].
    
    \bibitem{camci2018} U.~Camci, 
    Symmetry {\bf 10} (2018) no.12, 719.
	
	\bibitem{felix}
	M.~De Laurentis and A.~J.~Lopez-Revelles,
	Int.\ J.\ Geom.\ Meth.\ Mod.\ Phys.\  {\bf 11} (2014) 1450082
	[arXiv:1311.0206 [gr-qc]].
	
		
	\bibitem{Kofinas:2014owa}
	G.~Kofinas and E.~N.~Saridakis,
	Phys.\ Rev.\ D {\bf 90} (2014) 084044
	[arXiv:1404.2249 [gr-qc]].

	
	\bibitem{Bahamonde:2016kba}
	S.~Bahamonde and C.~G.~B\"ohmer,
	Eur.\ Phys.\ J.\ C {\bf 76} (2016) no.10,  578
	[arXiv:1606.05557 [gr-qc]].

	\bibitem{cristina}
	M.~De Laurentis, M.~Paolella and S.~Capozziello,
	Phys.\ Rev.\ D {\bf 91} (2015) no.8,  083531
	[arXiv:1503.04659 [gr-qc]].
	
	\bibitem{mauro}
	S.~Capozziello, M.~Francaviglia and A.~N.~Makarenko,
	Astrophys.\ Space Sci.\  {\bf 349} (2014) 603
	doi:10.1007/s10509-013-1653-5
	[arXiv:1304.5440 [gr-qc]].
	
	\bibitem{sergey}
	S.~Capozziello, M.~De Laurentis and S.~D.~Odintsov,
	Mod.\ Phys.\ Lett.\ A {\bf 29} (2014) no.30,  1450164
	[arXiv:1406.5652 [gr-qc]].
	
	\bibitem{micol}
	M.~Benetti, S.~Santos da Costa, S.~Capozziello, J.~S.~Alcaniz and M.~De Laurentis,
	Int.\ J.\ Mod.\ Phys.\ D {\bf 27} (2018) no.08,  1850084
	[arXiv:1803.00895 [gr-qc]].
	
	
	\bibitem{alcaniz}
	S.~Santos Da Costa, F.~V.~Roig, J.~S.~Alcaniz, S.~Capozziello, M.~De Laurentis and M.~Benetti,
	Class.\ Quant.\ Grav.\  {\bf 35} (2018) no.7,  075013
	[arXiv:1802.02572 [gr-qc]].

	
	
	\bibitem{Kofinas:2014aka}
	G.~Kofinas, G.~Leon and E.~N.~Saridakis,
	Class.\ Quant.\ Grav.\  {\bf 31} (2014) 175011
	[arXiv:1404.7100 [gr-qc]].
	
	\bibitem{Kofinas:2014daa}
	G.~Kofinas and E.~N.~Saridakis,
	Phys.\ Rev.\ D {\bf 90} (2014) 084045
	[arXiv:1408.0107 [gr-qc]].
	\bibitem{guzman}
	R.~Ferraro and M.~J.~Guzman,
	Phys.\ Rev.\ D {\bf 94} (2016) no.10,  104045
	doi:10.1103/PhysRevD.94.104045
	[arXiv:1609.06766 [gr-qc]].
	
	\bibitem{Capozziello:1996bi}
	S.~Capozziello, R.~De Ritis, C.~Rubano and P.~Scudellaro,
	Riv.\ Nuovo Cim.\  {\bf 19N4} (1996) 1.
	
	\bibitem{Capozziello:2012hm}
	S.~Capozziello, M.~De Laurentis and S.~D.~Odintsov,
	Eur.\ Phys.\ J.\ C {\bf 72} (2012) 2068
	[arXiv:1206.4842 [gr-qc]].
	
	\bibitem{kamen}
	A.~Y.~Kamenshchik, E.~O.~Pozdeeva, A.~Tronconi, G.~Venturi and S.~Y.~Vernov,
	Class.\ Quant.\ Grav.\  {\bf 31} (2014) 105003
	[arXiv:1312.3540 [hep-th]].
	
	\bibitem{Atazadeh:2011aa}
	K.~Atazadeh and F.~Darabi,
	Eur.\ Phys.\ J.\ C {\bf 72} (2012) 2016
	[arXiv:1112.2824 [physics.gen-ph]].
	H.~Dong, J.~Wang and X.~Meng,
	Eur.\ Phys.\ J.\ C {\bf 73} (2013) no.8,  2543
	[arXiv:1304.6587 [gr-qc]].
	B.~Tajahmad,
	Eur.\ Phys.\ J.\ C {\bf 77} (2017) no.4,  211
	[arXiv:1610.08099 [gr-qc]];  
	H.~Mohseni Sadjadi,
	Phys.\ Lett.\ B {\bf 718} (2012) 270,
	[arXiv:1210.0937 [gr-qc]].
	
	
	\bibitem{Capozziello:2016eaz}
	S.~Capozziello, M.~De Laurentis and K.~F.~Dialektopoulos,
	Eur.\ Phys.\ J.\ C {\bf 76} (2016) no.11,  629
	[arXiv:1609.09289 [gr-qc]].
	
	\bibitem{Kucukakca:2014vja}
	Y.~Kucukakca,
	Eur.\ Phys.\ J.\ C {\bf 74} (2014) no.10,  3086
	[arXiv:1407.1188 [gr-qc]];
	M.~Salti, O.~Aydogdu, H.~Yanar and F.~Binbay,
	Mod.\ Phys.\ Lett.\ A {\bf 32} (2017) no.34,  1750183;
	G.~Gecim and Y.~Kucukakca,
	arXiv:1708.07430 [gr-qc];
	S.~Bahamonde, U.~Camci, S.~Capozziello and M.~Jamil,
	Phys.\ Rev.\ D {\bf 94} (2016) no.8,  084042
	[arXiv:1608.03918 [gr-qc]].
	
	\bibitem{Bahamonde:2017sdo}
	S.~Bahamonde, S.~Capozziello and K.~F.~Dialektopoulos,
	Eur.\ Phys.\ J.\ C {\bf 77} (2017) no.11,  722
	[arXiv:1708.06310 [gr-qc]].
	
	\bibitem{horndeski}
	S.~Capozziello, K.~F.~Dialektopoulos and S.~V.~Sushkov,
	Eur.\ Phys.\ J.\ C {\bf 78} (2018) no.6,  447
	doi:10.1140/epjc/s10052-018-5939-1
	[arXiv:1803.01429 [gr-qc]].
	
	
	
	
	\bibitem{perivola}
	S.~Capozziello, S.~Nesseris and L.~Perivolaropoulos,
	JCAP {\bf 0712} (2007) 009
	doi:10.1088/1475-7516/2007/12/009
	[arXiv:0705.3586 [astro-ph]].
	
	
\end{thebibliography}
\end{document}